\documentclass[aps,nopacs,superscriptaddress,preprint]{revtex4}

\usepackage[english]{babel}
\usepackage[pdftex]{graphicx}
\usepackage{bm}
\usepackage{hyperref}
\usepackage{amsmath}
\usepackage[usenames]{color}
\usepackage{wrapfig}

\usepackage{ulem}

\advance\voffset -.25in

\usepackage{bm}
\newcommand{\cg}{} 

\begin{document}

\title{Predicting commuter flows in spatial networks using a radiation model based on 
temporal ranges} 

\author{Yihui Ren}
\affiliation{Physics Department and the Interdisciplinary Center for 
Network Science and Applications, University of Notre Dame, Notre Dame, IN 46556, USA}
\author{M\'aria Ercsey-Ravasz}
\affiliation{Faculty of Physics, Babes-Bolyai University, RO-400084 Cluj-Napoca, Romania}
\author{Pu Wang}
\affiliation{School of Traffic and Transportation Engineering, Central South University,
22 Shaoshan South Road, Changsha, 410075, Hunan, P.R.China }
\author{Marta C. Gonz\'alez}
\affiliation{Department of Civil and Environmental Engineering, Massachusetts Institute of 
Technology, 77 Massachusetts Avenue, Cambridge, Massachusetts 02139, USA}
\author{Zolt\'an Toroczkai}\email{toro@nd.edu}
\affiliation{Physics Department and the Interdisciplinary Center for 
Network Science and Applications, University of Notre Dame, Notre Dame, IN 46556, USA}

\date{\today}

\begin{abstract} 
Understanding network flows such as commuter traffic in large transportation networks is an ongoing challenge due to the complex nature of the transportation infrastructure and of human mobility.  Here we show a first-principles based method for traffic prediction using a cost based generalization of the radiation model for human mobility, coupled with a cost-minimizing algorithm for efficient distribution of the mobility fluxes through the network. Using US census and highway traffic data we show that traffic can efficiently and accurately be computed from a range-limited, network betweenness type calculation. The model based on travel time costs captures the lognormal distribution of the traffic and attains a high Pearson correlation coefficient (0.75) when compared to real traffic. Due to its principled nature, this method can inform many applications related to human mobility driven flows in spatial networks, ranging from transportation, through urban planning to mitigation of the effects of catastrophic events.
\end{abstract}

\maketitle

One of the  challenges in network science is predicting network flows from graph structural properties, node/edge attributes and dynamical rules. While for some networks (for example, electronic circuits) this is a well-understood problem, it is still open in general, and  especially for networks involving a social component \cite{Nature_BHG06,Gonzalez_Nature08} such as communication networks \cite{Krings2009,Onnela2007}, epidemic networks \cite{Eubank04,Colizza2006,Balcan2009} and infrastructure networks \cite{Hill06,Dorfler13,Motter13,Wilson1969,Makse1995,Barrett,Wu2006,Bono2010, Barthelemy_PR2011,Roth2011,WHBSG12,Ercsey12}. 
Here we focus on the  traffic flow prediction problem  in spatial networks, and in particular in roadway networks and validate our results using US highway network and traffic data  \cite{MITArcGIS}. Understanding flows in spatial networks driven by human mobility  would have many important consequences: it would enable us to connect throughput properties with demographic factors and network structure; it would inform urban planning  \cite{Krueckeberg1974,Batty08, BattyLongley94, Benenson04}; help forecast the spatio-temporal evolution of epidemic patterns \cite{Balcan2009,Eubank04,Colizza2006}, help assess  network vulnerabilities \cite{Holme02,PRE_ET12}, and allow the prediction of changes in the wake of catastrophic events \cite{Sept11}. 

When modeling transportation systems as networks we usually associate network nodes with locations and edges with physical paths between locations. Here, we define nodes as intersections between the roads and the road segment between two consecutive intersections as the edge connecting those nodes. We will refer to nodes also as sites or locations, interchangeably. Our ultimate goal is to determine the average traffic flow $T_{ij}$ expressing the number of flow units {\cg (for example vehicles)} per unit time (for example per day) through an edge $(i,j)$ of the network, given the network and the distribution of the {\cg population. 

For} any traffic to exist there must be people planning to travel between locations.  Given an origin location $a$ and destination $b$, the average number of travelers from $a$ to $b$ is determined by socio-demographic factors such as distribution of the population, availability of jobs, resource locations, etc. We define $\varPhi_{ab}$ as the average number of daily travelers planning to go from  site $a$ (origin) to  site $b$ (destination), where the average is computed over a longer time interval such as a year's period. We call $\varPhi_{ab}$ the mobility flux, or origin-destination (OD) flux and use the word {\cg flux} exclusively for that purpose. The socio-demographic model that describes the fluxes $ \varPhi_{ab}$ will be called  {\cg mobility law}. Note that the flux $\varPhi_{ab}$ does not tell us anything about the {\cg path chosen} between the origin and destination. It is simply the size of population at location $a$ planning to travel {\cg daily} to location {\cg $b$. When} people {\cg travel} from a location $a$ to a location $b$ they must choose a route on the network to do so. Accordingly, the $T_{ij}$ expresses the average number of daily travelers through edge $(i,j)$, which can in principle originate from any location $a$ traveling to any location $b$ as long as their chosen route on the network contains the road segment $(i,j)$. When referring to traffic on specific edges (road segments), that is the $T_{ij}$-s, we will use the word {\cg flow}, or traffic interchangeably. Note that $\varPhi_{ab}$ is well defined for any two nodes or locations $a$ and $b$ in the network, but it does not define any traffic (flow); whereas $T_{ij}$ is defined only for edges $(i,j)$ and it is a flow quantity. In analogy with physics $ \varPhi_{ab}$ corresponds to {\cg voltage}, whereas $T_{i,j}$ corresponds to {\cg current}.

{\cg Modeling} traffic flows in spatial networks can therefore be approached via solving two problems: 1) Determining the mobility fluxes $ \varPhi_{ab}$  for all origin-destination pairs  $(a,b)$  \cite{Simini2012,Zipf1946,Erlander1990}  and 2) Distributing the fluxes  $ \varPhi_{ab}$ through the network, that is determining the network paths along which the flow units are transported \cite{Wilson1969,Makse1995,Wu2006,Bono2010}. We call the first problem the Mobility Law problem and the second the Flux Distribution {\cg problem and present a solution to both problems in this paper.

The} common approach to the Mobility Law problem has been through the use of gravity models  \cite{Zipf1946,Wilson1969,Erlander1990,Jung2008,Krings2009,Kaluza2010,Barthelemy_PR2011}, which assume that the fluxes have the generic form $ \varPhi_{ab} = m_a^{\alpha} n_b^{\beta}/f(r_{ab})$  where $m_a$ and $n_b$ are the population sizes of  origin $a$ and destination  $b$, $r_{ab}$ is the distance between them, and $f(x)$ is called the deterrence function. Typical forms for $f$ are power-law $f(r_{ab})=r_{ab}^{\gamma}$ or exponential $f(r_{ab})=e^{d\,r_{ab}}$, where $\alpha$, $\beta$ $\gamma$ and $d$ are fitting parameters. As shown in Ref \cite{Simini2012} gravity models are essentially fitting forms and they have numerous ills. Besides not being based on first principles, the fitting parameters can vary wildly even within a single dataset (as function of $r_{ab}$) \cite{Jung2008,Kaluza2010,Krings2009,Viboud2006,Balcan2009}. They can also show non-physical behavior, for example when the destination has a large enough population, the number of travelers can exceed the size of the  origin {\cg population. Recently, } a novel mobility law  called the {\cg radiation model} was introduced using probabilistic arguments, which avoids the problems of gravity models \cite{Simini2012,Neda2013}. {\cg Here we will use the radiation model as the mobility law with a first-principles based generalization that allows us to couple it with the network structure, where mobility takes place.}

{\cg Given the $\varPhi_{ab}$  fluxes for all the $N(N-1)$ node-pairs $(a,b)$ obtained from the generalized mobility law, here we solve the Flux Distribution problem by using a cost-minimization principle, based on the expectation that commuters tend to minimize the cost of travel. This results in a novel, efficient capacity-aware flux distribution algorithm that helps predict traffic in roadway networks.}

\section*{\large Results}

\section*{A  {\cg cost based} radiation model}

\begin{figure}[htbp]
\centerline{\includegraphics[width = 6.0in ]{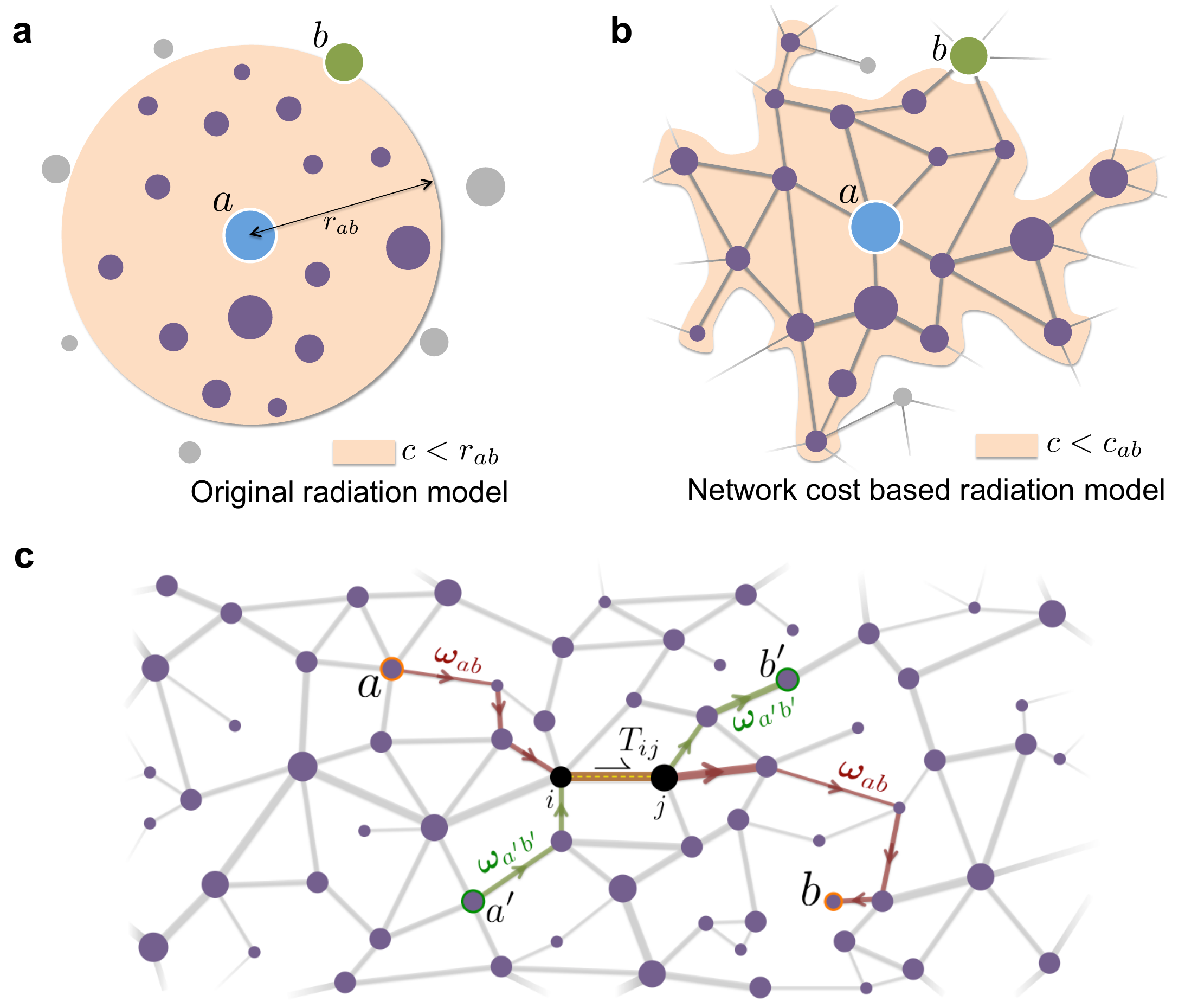}}
\caption{ {\bf  Schematics for traffic flow modeling.} ({\bf a}) The original radiation model uses distance $r_{ab}$ as a search criterion. ({\bf b}) The cost-based radiation model uses  network travel cost $c_{ab}$ as a search 
criterion, which usually has a heterogeneous distribution. ({\bf c}) The flow  $T_{ij}$ through edge $(i,j)$ is the sum of contributions from all those mobility fluxes $\varPhi_{ab}$ whose minimal cost paths $\omega_{ab}$ contain $(i,j)$.}\label{fig:1}
\end{figure}

The averaging in the definition of the flux $\varPhi_{ab}$ reduces the effect of fluctuations due to seasonal and occasional travel and thus it is expected to be determined mainly by travelers that commute regularly between home locations and job sites and regular freight traffic.  The radiation model is a socio-demographic model \cite{Simini2012} based on the assumption that people will search for the closest job opportunity that meets their expectation ({\cg see Supplementary Note 1}). The expectation of an individual is modeled by a single variable $z$ called the benefit variable, which acts as an absorption threshold: an individual ``emitted'' from location $a$ will take a job at another location $b$ (it {\cg becomes} {\cg absorbed} at $b$) only if the $z$ variable associated with the job site at $b$ surpasses that of the individual's and she could not find any such absorption site closer than $b$. Paper \cite{Simini2012} derives the expression of the probability $p_{ab}$ for an individual from location $a$ with population $m_a$ to find the closest job opportunity that meets her expectation at location $b$ with population $n_b$ and nowhere closer within a range of $r_{ab}$, where $r_{ab}$ is the distance between $a$ and $b$. Assuming independent emission-absorption events, the average mobility flux from $a$ to $b$ is then given by:
\begin{equation}
 \varPhi_{ab} = \zeta \,m_{a} p_{ab} = \zeta \, \frac{m_{a}^2 n_{b}}{(m_{a}+s_{ab})(m_{a}+s_{ab}+n_{b})}\;, \label{phiab}
\end{equation}
where $\zeta$ is the fraction of travelers in a location, considered to be an overall constant characterizing the whole of the population and $s_{ab}$ is the size of the population within a  disk of radius $r_{ab}$ centered on $a$, excluding the populations at locations $a$ and $b$, see Fig. \ref{fig:1}a.  
The distance $r_{ab}$ is interpreted as the crow flies, which, in heterogeneous environments does not usually correspond to the actual length of travel from $a$ to $b$. Here we extend the radiation model by saying that the individual will be choosing the site $b$ that has the lowest travel cost $c_{ab}$ {\cg on the network}, with a benefit factor at least as large as the individual's. We will refer to this model as the cost-based radiation model ({\cg see Supplementary Note 1}). We compare two travel cost measures, in particular one based on  path lengths ${\ell}_{ab}$ and the other based on travel times $\tau_{ab}$, both measured along roads. The path length ${\ell}_{ab}$ is the shortest distance (in km-s) from $a$ to $b$ along existing network paths, so it is closely related but larger than the geodesical radius $r_{ab}$ {\cg (measured as great-circle distance)}. The second travel cost {\cg measure} is the shortest time (in minutes) $\tau_{ab}$ it takes to go from $a$ to $b$ along the {\cg network }paths, and thus it depends on travel speeds as well. The expression for the fluxes is still given by (\ref{phiab}), however, the  population sizes $s_{ab}$ are computed differently.  Accordingly, the shape of the area {\cg around site $a$} with cost of travel not larger than $c_{ab}$ {\cg on the network} is no longer an annular disk with a dent as in Fig. \ref{fig:1}a, but it has an amoeboid shape as shown {\cg in Fig. \ref{fig:1}b}. There is an important difference between the criterion $r_{ab}$ used in ref \cite{Simini2012} and our general cost criterion $c_{ab}$. The former decouples the mobility law from the underlying transportation network, whereas the $c_{ab}$ (hence $s_{ab}$ and thus the $ \varPhi_{ab}$)  depend on the network of paths and their properties, thus coupling the mobility law with the network {\cg itself}.

\section*{Flux distribution without capacity limitation}

The total flow $T_{ij}$ through edge $(i,j)$ is generated by all those travelers that happen to have edge $(i,j)$ on a lowest cost path between their start and end locations. For a pair of origin-destination sites $(a,b)$, let us denote by $P_{ab}$ the set of all network paths from $a$ to $b$ and by  $\bm{\omega}_{ab} \in P_{ab}$ a minimal cost path.  Thus $\bm{\omega}_{ab}$ is a sequence of edges $\bm{\omega}_{ab} = \{(a,i_2),(i_2,i_3),\ldots,(i_L,b)\}$ such that $c_{ab} = \min_{\bm{\pi}_{ab}\in P_{ab}}\{ \sum_{(i_l,i_{l+1})\in \bm{\pi}_{ab}} c_{i_li_{l+1}} \}$ is attained for $\bm{\pi}_{ab}=\bm{\omega}_{ab}$ (see Fig. \ref{fig:1}c). Note that in principle, there might be several paths with the same lowest cost (called ``minimal'' paths hereafter) and this possible degeneracy must be included in the expression  of the total traffic flow through a given edge $(i,j)$ :
\begin{equation}
T_{ij} = \sum_{a,b\in V}  \frac{g_{ab}(i,j)}{g_{ab}} \,  \varPhi_{ab}\label{trf}\;.
\end{equation}
Here $g_{ab}$ is the number of minimal paths from $a$ to $b$ and $g_{ab}(i,j)$ is the number of minimal paths that contain edge $(i,j)$. When the cost $c_{ab}$ is not an integer value but a real number (physical distance or travel time), usually there is no degeneracy ($ g_{ab} = 1$ and $g_{ab}(i,j) = 1$ if $(i,j)$ belongs to $\bm{\omega}_{ab}$, zero otherwise) and (\ref{trf}) sums whole fluxes. According to (\ref{trf}), traffic values are obtained from sums of fluxes weighted by adimensional quantities, and thus traffic and flux have the same unit of measure. {\cg Realistic traffic data is typically provided in units of vehicles per day in which case we need to multiply the rhs of (\ref{trf}) with an overall constant representing the average number of vehicles per traveling person, here included into $\zeta$, for simplicity. Also for} simplicity, we will omit to indicate the unit of measure for fluxes and traffic, showing only numerical values, with the implicit assumption that they are in units of number of vehicles per day.

Eq. (\ref{trf}) is similar to the expression of edge betweenness centrality \cite{Newman_PRE01,Brandes2001, Sreenivasan2007,PRL_ET10,PRE_ET12},  with the difference being that instead of computing with the number of minimal paths, we now use weights of minimal paths, which are the mobility fluxes computed from the {\cg mobility} law (the cost-based radiation model in this case). Therefore, the flows $T_{ij}$ can be {\cg obtained} using the same algorithm as for weighted betweenness centrality  \cite{Newman_PRE01,Brandes2001,PRE_ET12} with two necessary modifications. 
\begin{figure}[htbp]
\centerline{\includegraphics[width = 5.8in ]{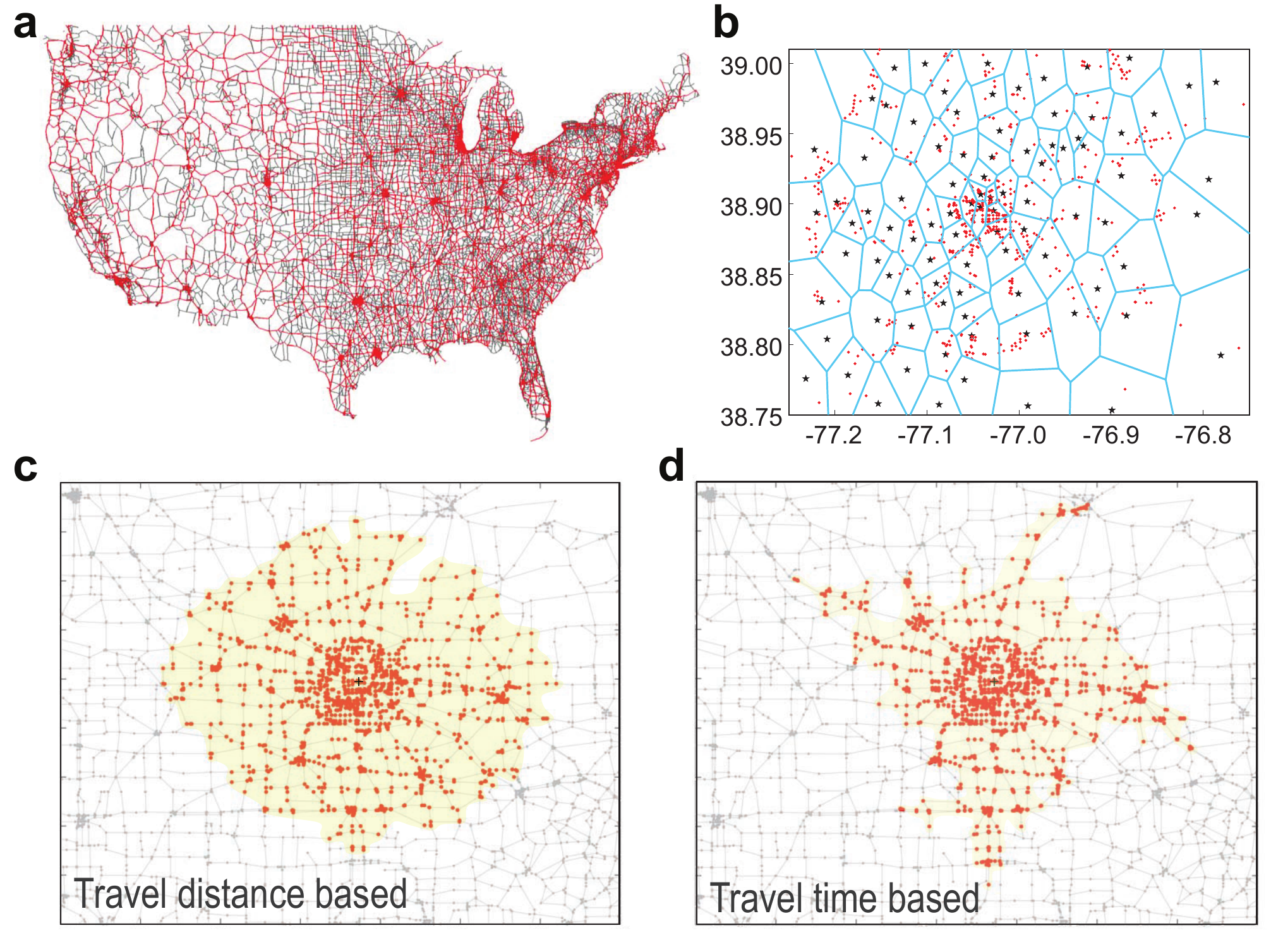}}
\caption{ {\bf Network and population data.} ({\bf a}) The US highway network with nodes as intersections and edges as road segments between intersections. It has $N = 137\,267$ nodes and $M = 174\,753$ edges. The red segments ($43\%$) have recorded annual average daily traffic values. ({\bf b}) Assigning a population size {\cg (see the Methods section)} to every intersection (red dots) using a Voronoi mesh and  zip-code level census data (zip-code centers indicated by black stars); Washington DC area is shown. ({\bf c}) Geographical area of locations around a node centered in Minneapolis, MN, with travel cost $c_{ab}$ not larger than a  given value using travel distance $\ell_{ab}$ as travel cost. ({\bf d}) Same as ({\bf c}), but using travel time $\tau_{ab}$ as travel cost. }\label{fig:2}
\end{figure}
One concerns implementation (see Methods section) and the other exploits the notion of range-limitation. For realistic size networks (infrastructure networks with hundreds of thousands to millions of nodes) the computation of (\ref{trf}) for all edges can become unfeasible (especially for collecting statistics). One can reduce the computational costs by introducing a range-limit on how far (in cost measure) we build the minimal paths  tree (MPT) from the source (root) node \cite{PRE_ET12,PRL_ET10}. In particular we only build the largest MPT from root  $a$ such that for all nodes $v$ in it we have $c_{av} \leq C$. The rationale is that beyond a cost threshold $C$ the contribution of the corresponding mobility fluxes is very small. The full-range algorithm has a complexity of ${\cal O}(N M \log N)$, where $N$ is the number of nodes and $M$ is the number of edges. In the case of US Highways (sparse network) this is a computation on the order of $10^{10}-10^{12}$, which is relatively costly. However, as we show in later sections, for the case of contiguous US, range limitation can reduce this complexity by several orders of magnitude without considerably affecting the accuracy of the results. 

\section*{Flux distribution with capacity limitation}

Network congestion is a ubiquitous phenomenon, resulting from edges having a finite transmission capacity.  We define the transmission capacity $C_{ij}$ of { an edge} $(i,j)$ as the largest daily flow value above which individuals 
will choose alternative routes with high probability.  Next we show how to distribute the mobility fluxes in a capacity limited network assuming that all the  $C_{ij}$ values are known. 

We use dynamic distribution of the traffic by gradually increasing the number of travelers until the first $q$ congested edges appear. The congested edges are then removed from the network for further traffic. More travelers are subsequently added to the network until another { $q$} edges become congested, which are then closed for further traffic, and this process is repeated until all travelers have been distributed into the network. Ideally ${ q} = 1$, but it is better to choose $ q > 1$ (such as $q = 100$, but still with ${ q} \ll M$), because on one hand congestion thresholds in finite systems are not sharp and thus $q > 1$ serves as a ``softness'' parameter, and on the other hand it speeds up the computations.  

Let us denote by $t_{ij}(G)$ the flow on the edges of a {\cg network (or graph)} $G$ computed using Eq.  (\ref{phiab}) with $\zeta = 1$, that is with  $ \varPhi_{ab} = m_a\,p_{ab}$. Note that the multiplicative coefficient $\zeta$ in the mobility fluxes (\ref{phiab}) is also multiplicative in the { traffic (or flow)} values.  Let us denote by $G_{n}$ the graph obtained from $G_{n-1}$ after removing the set $L_n$ of $q$ congested edges in the $n$-th step. We define recursively $\zeta_n =  \left\langle C_{ij;n-1}/t_{ij;n} \right\rangle_{L_n}$ with $T_{ij;0} \equiv 0$, $G_0 = G$, where $t_{ij;n} = \prod_{r=1}^{n-1}(1-\zeta_r)t_{ij}(G_{n-1})$ is the non-adjusted traffic coming from mobility fluxes $ \varPhi_{ab}$ corresponding to the fraction of the population not already in the network in that step and $C_{ij;n-1} = C_{ij} - T_{ij;n-1}$ are the corresponding reduced capacities in $G_{n}$. The set $L_n$ is defined as the $q$ edges with the smallest ratios $C_{ij;{n-1}}/t_{ij;n}$.  In the Methods section we show that after $k$ iterations the final flow becomes:
\begin{equation}
T_{ij;k} = \alpha_1 t_{i,j}(G) + \alpha_2  t_{i,j}(G_1)+ 
\ldots + \alpha_k  t_{i,j}(G_{k-1})  \label{tijk}
\end{equation}
where
\begin{equation}
\alpha_n = \left\langle \frac{C_{ij;n-1}}{t_{ij}(G_{n-1})} 
\right\rangle_{L_n} = \zeta_{n}  \prod_{r=1}^{n-1}(1-\zeta_r)\;,  \;\;\;
n=1,\ldots,k\,.
\end{equation}
The total number of iterations $k$ (stopping criterion) is determined by having all the traveling population $\zeta \sum_i m_i = \zeta m$ distributed onto the network, that is, $k$ is the  smallest integer for which
\begin{equation}
\alpha_1 + \alpha_2+\ldots+\alpha_k \geq \zeta \label{suma}
\end{equation}
{\cg holds.}

\section*{Comparison with empirical data}

To validate our approach we compared the model's output with real traffic data from a US highway network database \cite{MITArcGIS}, which consists of $M=174\,753$ road segments (edges) and $N = 137\,267$ intersections (nodes). The node features are longitude and latitude and the edge features are the IDs of the end nodes, road length, road class, number of lanes and annual average daily traffic (number of vehicles per day). The traffic values are available for about 43\% of all edges (road segments)  randomly distributed throughout the continental US (see Fig. \ref{fig:2}a) providing a good statistical basis for comparisons. 

Traffic values were generated for all road segments by the model via Eqs (\ref{trf}) or (\ref{tijk}-\ref{suma}) following the methods described in the previous sections (also see {\cg Supplementary Method 1}). The computation of the fluxes $\varPhi_{ab}$ for all origin-destination pairs requires the knowledge of the population sizes at the level of intersections (nodes). To that end, population sizes at the level of intersections were generated using population data from the US Federal Zip Code database \cite{zipcodes} and a Voronoi mesh based partitioning (Fig. \ref{fig:2}b) as described in the {\cg Methods section}. 

We compare two statistical quantities between the model output and data. One is the overall distribution of traffic flow values (specifically the logarithm of the traffic, justified below) and the other is the Pearson correlation coefficient (PCC) between the predicted traffic flow and the actual traffic flow on the edges where this data is available. Note that the PCC is computed not with logarithmic traffic  values but actual traffic values. The PCC is a much more stringent comparison criterion as it tests for the strength of linear relationship between model and data. The higher the PCC, the higher the ability of the model to predict traffic flow values at the individual edge (road segment) level. 

As discussed in the paragraph under Eq. (\ref{trf}) the rather costly computation of the traffic using Eqs (\ref{trf})-(\ref{suma}) can be performed efficiently if we include only those origin-destination fluxes $\varPhi_{ab}$ for which the travel cost $c_{ab}$ is below some threshold (range limitation). Before we compare the traffic values, in the next section we show that the mobility fluxes obey a simple scaling law over several orders of magnitude, which then can be exploited to determine the range limit for accurate and efficient traffic computations.

\subsection*{{\cg A scaling law for the mobility fluxes in the contiguous US}} 

\begin{figure}[htbp]
\centerline{\includegraphics[width = 5.8in ]{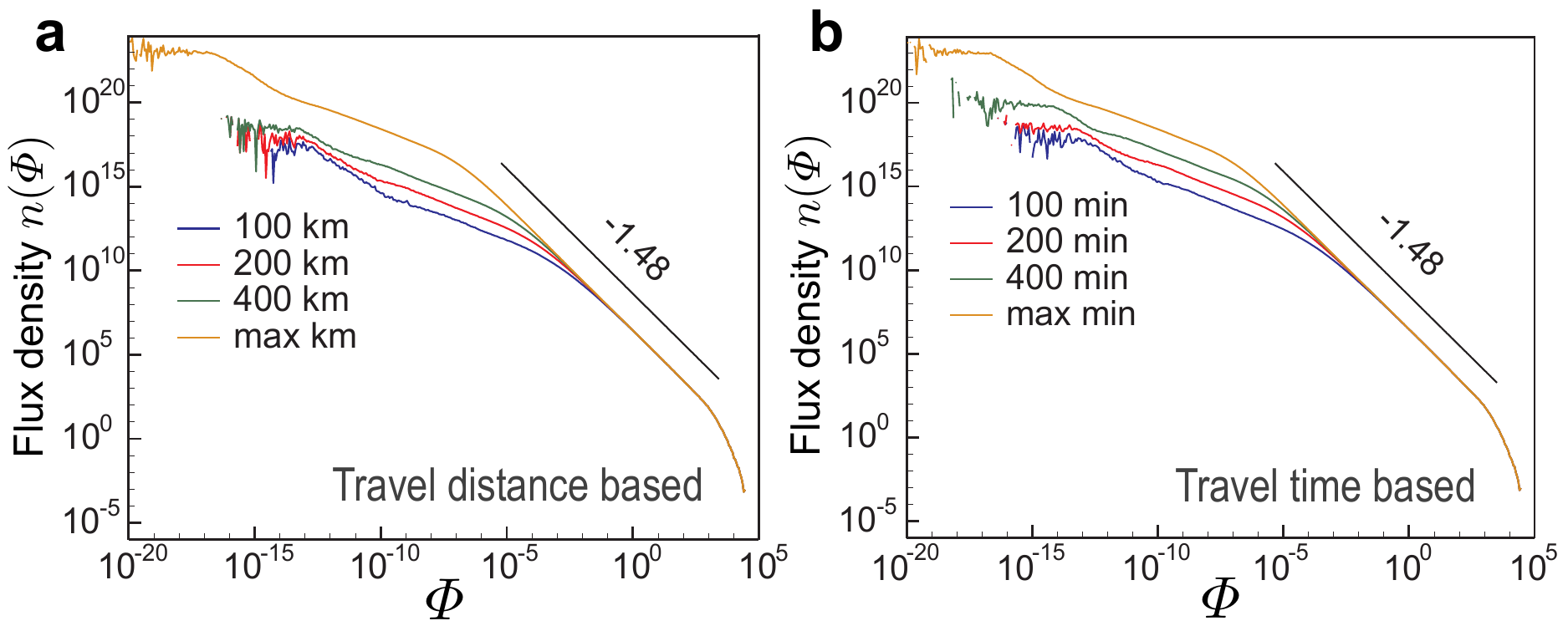}}
\caption{ {\bf  Mobility fluxes, a scaling law.} ({\bf a},{\bf b}) Density of { origin}-destination pairs with mobility flux  $\varPhi$ (with $\zeta=1)$ based on ({\bf a}) travel distance cost function $c_{ab} = \ell_{ab}$ and ({\bf b}) travel time cost function $c_{ab} = \tau_{ab}$.  The orange curve corresponds to no range limitation, namely, it includes all origin-destination pairs. Using a travel distance based range limit of $100 \mbox{km}$ or more ({\bf a}), or of travel time based limit of $100 \mbox{min}$ or more ({\bf b}) all curves collapse in the range of significant flux values.}\label{fig:3}
\end{figure}

Using the distribution of the population and the roadway network  from the data we computed the $ \varPhi_{ab}$ mobility fluxes via the cost-based radiation model (\ref{phiab}), using both travel distance $\ell_{ab}$ and travel time  $\tau_{ab}$ as travel cost, to determine $s_{ab}$ {\cg (Supplementary Method 1)}. Let  $n(\varPhi)$ denote the un-normalized number density of origin-destination (OD) pairs with mobility flux $\varPhi$, that is $d \varPhi \, n( \varPhi)$ is the number of OD pairs with fluxes in the range $[ \varPhi, \varPhi+d \varPhi)$ and $\int d \varPhi \, n( \varPhi) = \zeta \sum_i m_i$ is the total flux. Figs. \ref{fig:3}a,b show that the mobility flux density follows a power-law 
\begin{equation}
n( \varPhi) \sim  \varPhi^{-\mu}\;,\;\;\;\mu \simeq 1.48 \label{scaling}
\end{equation}
holding for over seven orders of magnitude.  Note that it actually holds for over nine orders of magnitude, however, we may neglect the very small flux values (below $10^{-4}$) as they do not contribute significantly to traffic. The scaling behavior (\ref{scaling}) can be derived from a counting argument using (\ref{phiab}), described {\cg as follows.} At intermediate to large ranges for $c_{ab}$, the population $s_{ab}$ within the ameboid domain is much larger than those at sites $a$ or $b$: $s_{ab} \gg \max (m_a,n_b)$ and therefore $ \varPhi_{ab} \simeq \zeta m_a^2 n_b s_{ab}^{-2}$. Assuming a typical population size $\langle m \rangle$ at any node, {\cg we have} $\varPhi_{ab} \simeq \zeta \langle m \rangle k_{ab}^{-2}$, where $k_{ab}$ is the number of nodes within the ameboid domain. Moreover, $k=k_{ab}$ is also the index of the node on the minimal path tree  centered on $a$ (index 0) just before node $b$. Since the index has a uniform distribution, we can use the method of inverse transform: $ \varPhi'(k) = -2\zeta \langle m \rangle k^{-3}$, $k_0( \varPhi) = (\zeta \langle m \rangle)^{1/2}  \varPhi^{-1/2}$ so $n( \varPhi) = 1/ | \varPhi'(k_0)| = \frac{1}{2} (\zeta \langle m \rangle)^{1/2}   \varPhi^{-3/2}$, and thus $\mu = 3/2$. In {\cg Supplementary Note 2} we show that an alternative approach using the assumption $s_{ab} \sim c_{ab}^2$ and computing thus the distribution of $ \varPhi_{ab} \sim c_{ab}^{-4}$ while also leads to a power-law, it generates an exponent of $1.3$ ({\cg Supplementary Figure 3}). The reason for why this approach generates a different exponent for the flux distribution is because the assumption $s_{ab} \sim c_{ab}^2$ does not hold for the roadway network due to the fractal-like nature \cite{BattyLongley94, MHS95,Batty08} of the ameboid domains; instead it obeys a scaling $s_{ab} \sim c_{ab}^{\nu}$ with $\nu \simeq 1.33$ ({\cg Supplementary Figure 4}). This observation provides additional support to studies of the fractal morphology and the underlying roadway networks of urban sprawls \cite{Batty08, Bettencourt07}.

The scaling law (\ref{scaling}) implies that over several orders of magnitude the origin-destination fluxes are heterogeneous and scale-invariant, namely, fluxes from fractional values to hundreds of thousands of vehicles are transported across the highway network, daily. This, in turn determines the width of the traffic distribution which, as shown in the following sections, obeys a lognormal  distribution. The power-law (\ref{scaling}) is a consequence of the scaling $\varPhi_{ab} \sim  s_{ab}^{-2}$, which in turn is a consequence of the threshold condition for mobility in the radiation law {\cg (Supplementary Note 1)} that is, of the fact that individuals will travel to the site that meets their expectation and it is the least costly to reach on the network. 

\subsection*{Network flow modeling}

The traffic values were computed on all edges using Eqs. (\ref{trf}-\ref{suma}) and compared to real traffic values on the subset of edges  for which this data is available (red edges in Fig. \ref{fig:2}a). Fig. 4 shows the comparisons using the density  of log traffic $\rho(\log_{10}(T))$ and the PCCs between data and model traffic values.   
\begin{figure}[htbp]
\centerline{\includegraphics[width = 3.8in ]{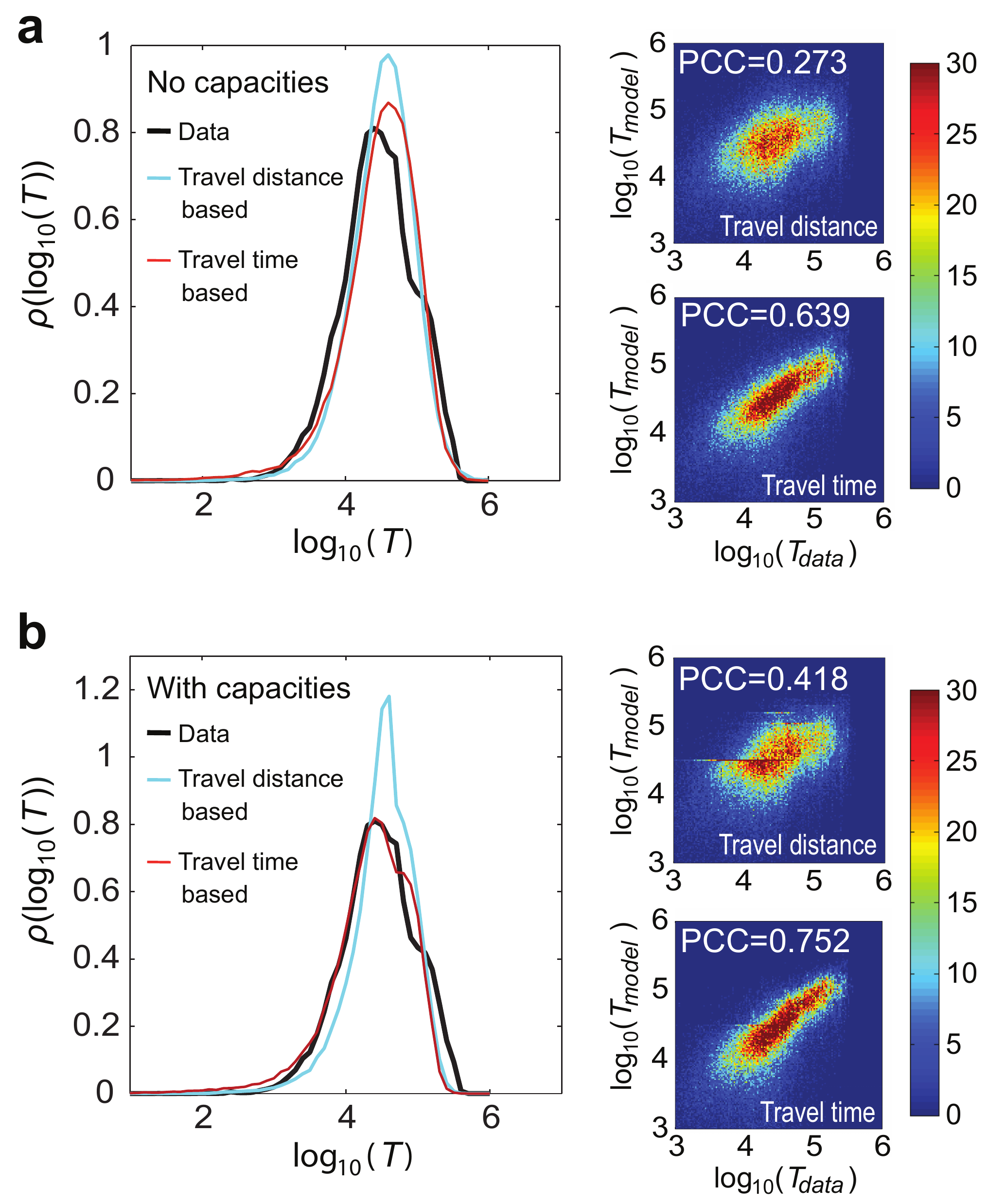}}
\caption{ {\bf Comparison with data.} ({\bf a}) Left panel: comparison between the densities of log(traffic) obtained from data (black line) and the model without capacity limitation based on travel distance (blue line) and travel time (red line). The heat maps  (right panels) are the scatter plot between real log traffic and model log traffic values without capacity limitation.  The linear bin size is 0.02 in the heat maps and the color bar gives the number of events (road segments) that fall within the same bin. For the upper map the travel distance cost function (with a range limit of 400 km) was used, generating a PCC of 0.273. For the lower map the travel time cost function was used with a range limit of 400 min and velocity classes 90-40-15 mph ({\cg Supplementary Method 1}), generating a PCC of 0.639.  ({\bf b}) is similar to  ({\bf a}) but with capacity limitation ({\cg Supplementary Method 2}).  Here the PCC of 0.752 was obtained with the same velocity class configuration as in ({\bf a}). The range limits were 100 km and  100 min, respectively. For the computation with capacity limitation and time costs, the iterations were stopped when Eq. (\ref{suma}) was first satisfied, after 83 iterations corresponding to about $2.37\%$ of the edges being congested. } \label{fig:4}
\end{figure}
The case without capacity limitation is shown in Fig. \ref{fig:4}a. The overall multiplying factor $\zeta$ in the model was set to match the mean of the distribution of traffic in the model with that in the data. As shown in the left panel of Fig. \ref{fig:4}a, the model distributions (blue and red lines) track rather closely the log traffic distribution (black line) of the data with a slightly better agreement when using travel time based cost functions. The PCC-s, however, show a significant difference, 0.273 vs 0.639, indicating that travel time is a much better criterion for evaluating cost of travel than travel distance.  Although for the travel distance based model there are no other adjustable parameters, one could state that for the travel time based case, however, the velocities provide enough wiggle room to achieve the much better fit with the data. While indeed, the fit is improved by varying the velocities, this is not the main reason for the agreement. The typical  travel velocities  were obtained using a consistent procedure described in {\cg Supplementary Method 1}. In order to avoid too many fitting parameters, we have not used separate velocities for individual roads, but all roads were lumped into three velocity ranks: fast, medium and  within-city speeds.  For the velocity combinations tested shown in {\cg Supplementary Table I}, the corresponding PCCs  were all found to be above 0.61, still much higher than the 0.27 PCC from the travel distance based model. 
 \begin{figure}[htbp]
\centerline{\includegraphics[width = 5.90in ]{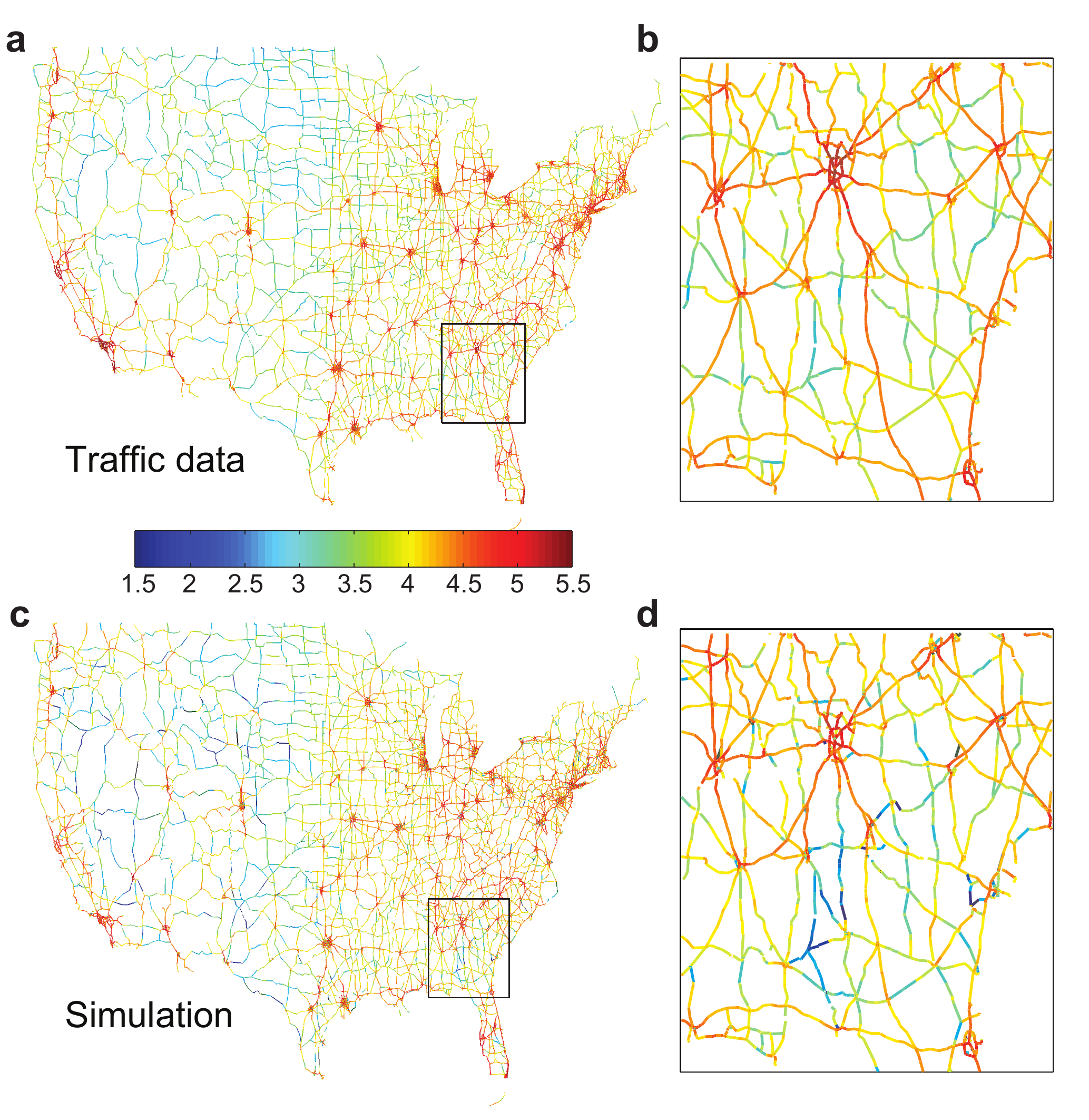}}
\caption{{\bf A visual comparison.} ({\bf a}) {\cg Log} traffic values indicated via colors (see color bar) on major highways in the contiguous US. ({\bf b}) Magnification of a south-east region. ({\bf c}) Same as in ({\bf a}) {\cg but for the model output} using travel time cost with capacity limitation and with the same parameters as in Fig \ref{fig:4}b. ({\bf d}) magnification of the same region from ({\bf c}) as in ({\bf a}). }\label{fig:5}
\end{figure}
A better agreement can be achieved if capacity limitation is taken into consideration ({\cg Supplementary Method 2}), see Fig \ref{fig:4}b. The distributions of the log traffic show an even better match, and the highest obtained PCC is 0.752 when using travel time costs. In the case of capacity limitation, the iterations were stopped when condition (\ref{suma}) was satisfied.  Fig. \ref{fig:5} shows roadway traffic values (using colors to indicate the volume of the traffic) for visual comparison between model and data, showing a relatively good agreement between the two, for most of the roads.
 
\begin{figure}[htbp]
\centerline{\includegraphics[width = 3.8in ]{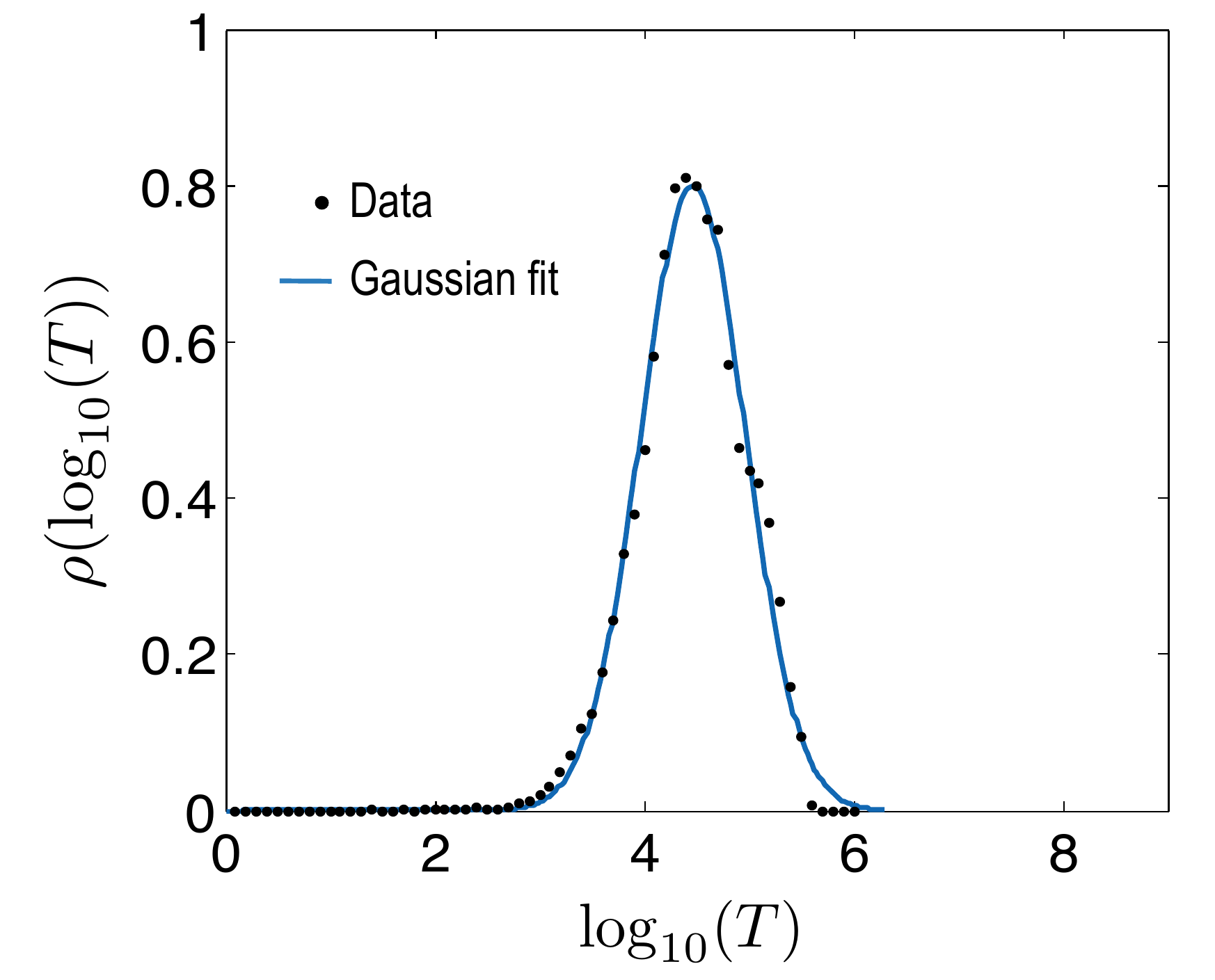}}
\caption{{\bf Distribution of traffic.} The real (data) traffic distribution is well approximated by a log-normal. }\label{fig:6}
\end{figure}

The traffic values were generated using the weighted betweenness centrality type expression (\ref{trf}). Based on this we can give an analytic argument for why the shape of the traffic density plotted in Fig. \ref{fig:6} is lognormal. 
It was previously shown   \cite{PRL_ET10,PRE_ET12} that (for example Eq. (6) of Ref  \cite{PRL_ET10}) the natural scaling variable for the betweenness distribution is the logarithm of the betweenness (hence traffic) and that the betweenness distribution can be written as a convolution between the degree distribution $P(k)$ and the distribution function $\varPsi_r$ of the deviation ({\cg noise}) of the shell sizes  (the number of network nodes at a given range $r$) from its scaling form described by the corresponding branching process characteristic for that network class. That is, if $b$ denotes the betweenness variable, $p(b) \sim \frac{1}{b} \int dk \, P(k) \varPsi_r (\log b - \log \beta_r - \log k)$. For spatial networks such as random geometric graphs, or roadways, this scaling form is power-law with the exponent given by the dimensionality of the embedding space ($d=2)$ that is $\beta_r  \sim r^d = r^2$.  Since our degree distribution is almost uniform we can replace $P(k) \sim \delta (k-\langle k \rangle)$ with good approximation, which from above leads to $p(b) \sim \frac{1}{b} \varPsi_r (\log b - \log \beta_r - \log \langle k\rangle)$. As shown in Refs \cite{PRL_ET10,PRE_ET12} $\varPsi_r$ is Gaussian for large random networks (also for the US highway network), and thus the betweenness/traffic distribution becomes a log-normal, indeed supported by Fig \ref{fig:6}.

\section*{\large Discussion}

There are several gravity models in the literature that may be used to better match the local  traffic, but they come at the expense of additional fitting parameters  {\cite{Jung2008,Kaluza2010,Krings2009,Viboud2006,Balcan2009}}. However, if we would need to predict new flow patterns in the wake of network changes (for example due to natural disasters) it is not clear  what values should be used for the fitting parameters on the changed network. The main strength of our approach is that it is based on first principles  and thus it can be easily used for flow predictions in the wake of network changes.   The model can be further improved by adding more features such as a better approximation to  population distribution at the intersection level, seasonal variations, etc. And indeed, we have seen the agreement improving already by including capacity limitations, even with crude approximations for travel speeds. At every step, our modeling approach follows the Maximum Entropy Principle by Jaynes \cite{Jaynes57} in the sense that the model incorporates only known data (population distribution, the network and capacities) and the assumed behavior (cost-based radiation law and cost minimizing path-choice); for everything else it assumes uniform distributions with minimum parameters so as to minimise biases (such as the coefficient $\zeta$ or the distributions within speed categories).  

{\cg The original radiation model treats costs simply as a geometric range, it does not involve any transportation network. Since our framework allows the use of any cost-function, we could  still use the original radiation model for calculating the fluxes $\varPhi_{ab}$ by calculating  the area populations $s_{ab}$ using geodesic, or in this case, great-circle distances. However, we cannot use great-circle distances to find the lowest travel cost paths on the network because great-circle distances say nothing about network paths. Thus, we would be forced to employ two, somewhat inconsistent travel cost criteria: when estimating the area population that we can reach ($s_{ab}$) we would use as-crow-flies distances, but when computing network paths for travel, we have to revert to network-based travel costs. This would lead to errors in geographically heterogeneous areas, where a direct path to a location may run through an obstacle (such as a lake, a mountain, a gorge, etc.), and thus that location would be included into $s_{ab}$, but the real network path would avoid the obstacle at a more significant cost (excluding that location from $s_{ab}$). Statistically, however, using the original radiation model would not lead to large errors in the traffic distribution $\rho$ and the PCC for a large country as US. The reason is because using great circle distances we still get a good approximation of the population $s_{ab}$ on the network for most OD pairs $(a, b)$. Both the PCC and the traffic distribution imply sums/averages taken over a large fraction of the whole US and these averages are dominated by short and medium distances, which are abundant in heavily populated areas. With some exceptions, heavily populated areas tend to be in regions where mobility is not hampered by geographical obstacles and thus in these heavily populated areas network paths tend to run in the direction of the shortest geometrical distance, making the two cost measures proportional to one another.

Besides consistency, our model also has a computational advantage in that we can simultaneously find the lowest cost paths and the population values $s_{ab}$ (within the Dijkstra part of the algorithm, see the Methods section), within one run of the algorithm. However, when computing the fluxes $\varPhi_{ab}$ using great-circle distances we need a separate algorithm of an entirely different nature, which is in addition to the flux distribution code. This additional algorithm needs to find all the points increasingly by their great-circle distance from an origin $a$, then it needs to do this for all  ($N$) origins. This is a well-known problem in computational geometry and the most efficient implementation runs in $O(N^2 \log N )$ time \cite{Dickerson}. Thus, since the flux distribution algorithm is also of $O(N^2 \log N )$ complexity (the roadway network is sparse), this additional algorithm essentially doubles the computational time (confirmed by our simulations). }

In summary, the cost-based radiation model provides a feasible approach to model flows in spatial networks where the choice of transport paths on the network is driven by a cost-minimization principle, given the distribution of population and resources. The mobility fluxes are generated by the individuals finding those absorption sites on the network that meet their expectation thresholds and that are the least costly to reach on the network. This couples the socio-demographic aspect (Mobility Law) with the network transport aspect (Flux Distribution)  and the final flow will be the result of the interplay between these two aspects.  Due to its principled nature, we expect that the modeling approach presented here is applicable with some modifications not just for highway network datasets but for spatial networks in general where traffic is generated by a cost-incurring transport. 

\section*{\large Methods} \label{Methods}

{\cg \section*{Assigning populations to network vertices}

 In order to compute the mobility fluxes $\varPhi_{ab}$ we need to know not only the populations at sites $a$ and $b$ but also at all sites around $a$ within the domain defined by the cost function $c_{ab}$. As we are modeling traffic at the level of road intersections, we need to resolve the distribution of population at this level. For this purpose, we used population information from the US government's zipcode database \cite{zipcodes}. Restricted to the contiguous US, the corresponding population data came from 31\,343 zip code instances. However, there are $N = 137\,267$ network vertices (intersections), which implies that a finer resolution is needed than what is provided by zip codes, for population. We perform this refinement  in two steps. First, we construct a 2D Voronoi diagram using the set of points (Voronoi sites) provided to us in the zip code data (these usually correspond to post-office locations, given in (long, lat)) and assign every intersection (network node) to that Voronoi site to which it is the closest.  Second, we label those Voronoi cells that had no intersections assigned to them (26\%). We remove their sites temporarily, then we redo the Voronoi mesh with these labeled sites absent. Next we place back the labeled sites and find those Voronoi cells from the second mesh that contain these labeled sites. We then add the population of the labeled sites to the population of those cells from the second mesh that contain them, and redistribute the population amongst the intersections within all cells of the second mesh, uniformly, see Fig. \ref{fig:2}b. This way no population is lost and they are all assigned naturally to the closest intersections.}

\section*{Weighted betweenness centrality algorithm}

This algorithm  proceeds by constructing the minimal paths tree (MPT) rooted at a vertex $a$, for all vertices $a$ using Dijkstra's algorithm \cite{Dijkstra1959}  (based on breadth-first search). Then starting from the leafs (the furthest nodes from the root $a$) of the MPT  it computes recursively for every edge $(i,j)$ of the MPT the contributions in the sum (\ref{trf}) coming from all paths with source node  $a$. Note that for a given root (source) node $a$ only those fluxes $ \varPhi_{av}$ contribute to these sums for which $v$ is part of the corresponding MPT. Thus, we don't need to generate all the fluxes $ \varPhi_{ab}$ for all pairs beforehand (which would be on the order of $2\times10^{10}$ values for the US highway system), but we can compute them locally when generating the minimum paths tree. 

\section*{Distributing flows in networks with capacity limitation} 

Consider the first step $k = 1$. {\cg Denoting the whole graph with $G$ and its edge set by $E$,} starting with $G$ we compute the non-adjusted flow values $t_{ij;1} \equiv t_{ij}(G)$ on all edges. We identify the set $L_1$ of { $q$} roads with the smallest  $C_{ij}/t_{ij;1}$ ratio, which are the roads that become congested early on. Define:  
\begin{equation}
\zeta_1 = \left\langle \frac{C_{ij}}{t_{ij;1}} \right\rangle_{L_1}\;,
\end{equation}  
where $\langle \cdot \rangle_{L_1}$ is an average taken over the edges in $L_1$. For edges in $L_1$, $\zeta_1 t_{ij;1}$ will be near their capacity $C_{ij}$ (if { $q$} is not too large). This allows for fluctuations around the congestion capacities, modeling the softness effect mentioned in the main text. The adjusted flow on edge $(i,j)$ at the end of the first step will therefore be 
\begin{equation}
T_{ij;1} = \zeta_1 \, t_{ij;1}\;, \;\;\; \forall (i,j) \in G.
\end{equation}
On the non-congested edges  $(i,j) \not \in L_1$, the new capacity will be $C_{ij;1} = C_{ij} - T_{ij;1}$.   In the next step ($k=2$) we consider the new graph $G_1$ with edge-set $E_1 = E \setminus L_1$ (removed the $q$ congested edges identified in the previous step). We then compute the non-adjusted flow $t_{ij;2} = (1-\zeta_1)t_{ij}(G_1)$ for all edges of $G_1$. The latter corresponds to flow computed with mobility fluxes $ \varPhi_{ab} = (1-\zeta_1) m_a p_{ab}$ because a $\zeta_1$ fraction of the population is already on the roads. We now identify the set $L_2 \subset E$  of $q$ edges ({ $|L_2| = q$}) with the smallest ratios $C_{ij;1}/t_{ij;2}$ and define:
\begin{equation}
\zeta_2 = \left\langle \frac{C_{ij;1}}{t_{ij;2}} \right\rangle_{L_2} = 
\frac{1}{1-\zeta_1} \left\langle \frac{C_{ij} - T_{ij;1}}{t_{ij}(G_1)} \right\rangle_{L_2}\;.
\end{equation}  
Then, the new, adjusted flow on the edges of $G_1$ will be 
\begin{equation}
T_{ij;2} =  T_{ij;1} + \zeta_2 t_{ij;2}  =  \zeta_1 t_{ij}(G) + (1-\zeta_1) \zeta_2 t_{ij}(G_1)\;,
\end{equation}
$\forall (i,j) \in G_1$, with the new capacities for further traffic becoming $C_{ij;2} = C_{ij} - T_{ij;2}$. In the third step $k=3$, we compute the non-adjusted flow $t_{ij;3} = (1-\zeta_1)(1-\zeta_2) t_{ij}(G_2)$, from fluxes $ \varPhi_{ab}=
(1-\zeta_1)(1-\zeta_2) m_a p_{ab}$ corresponding to the fraction of population not in the network, where $G_2$ is obtained from $G_1$ by removing the edges in $L_2$. We then identify the set $L_3$ of { $q$} edges with the smallest $C_{ij;2}/t_{ij;3}$ ratios and compute:
\begin{equation}
\zeta_3 = \left\langle \frac{C_{ij;2}}{t_{ij;3}} \right\rangle_{L_3} = \frac{1}{(1-\zeta_1)(1-\zeta_2)} \left\langle \frac
{C_{ij} - T_{ij;2}}{t_{ij}(G_2)} \right\rangle_{L_3}
\end{equation} 
yielding  the adjusted flow on all the roads $(i,j)$ of $G_2$:
\begin{equation}
T_{ij;3} =  T_{ij;2} + \zeta_3 t_{ij;3}  = \alpha_1 t_{ij}(G) + \alpha_2 t_{ij}(G_1) + \alpha_3 t_{ij}(G_2), \label{Tij3}
\end{equation}
$\forall\,(i,j) \in G_2$, where $\alpha_1 = \zeta_1$, $\alpha_2 = \zeta_2(1-\zeta_1)$, $\alpha_3=\zeta_3(1-\zeta_2)(1-\zeta_1)$. Thus, in the first step we distributed $\zeta_1 m = \alpha_1 m$ travelers, in the second step another $(1-\zeta_1)\zeta_2 m = \alpha_2 m$, in the third $(1-\zeta_1)(1-\zeta_2)\zeta_3 m = \alpha_3 m$, etc. A straightforward generalization of this yields the equations in the main text. 

{\cg \section*{Determining the effective range-limitation}

The very small mobility flux values in Fig. \ref{fig:3}a,b are coming from origin destination pairs whose separation involves a large travel cost $c_{ab}$. However, we expect that fluxes that are too small ($10^{-4}$ and smaller) do not 
contribute significantly to any traffic flow value, implying that we may limit our computaton of fluxes to ranges that generate fluxes that are not too small. To assess when range limitation is effective we have computed the fraction of population from a location $a$ traveling to sites whose travel cost (from $a$) is beyond a given threshold value $c_{ab} = R$:
$\epsilon_{a} = \sum_{b} (\varPhi_{ab}  - \varPhi_{ab}^{R})/\sum_{b} \varPhi_{ab}  = 1-  \sum_{b} p_{ab}^{R}$, where  $\varPhi_{ab}^{R} = \varPhi_{ab}$ if $c_{ab} \leq R$ and zero otherwise, and  we used the expressions $\varPhi_{ab} = \zeta m_a p_{ab}$ and $\varPhi_{ab}^{R} = \zeta m_a p_{ab}^{R}$.  This fraction $\epsilon_{a}$ is the probability that a person from location $a$ will travel beyond range $R$, which is then omitted from traffic flow calculations with range limit $R$. {\cg Supplementary Fig. 6a,b} shows the cumulative fraction of the locations with long-range (larger than $R)$ travel probability less than $\epsilon$. When cost of travel is computed based on travel distance, we see that for 95\% of all locations the likelihood of daily long-range travel is less than $1\%$, $0.2\%$ and $0.05\%$ when going beyond $100$ km, $200$ km and $400$ km respectively. In terms of travel time cost,  95\% of all locations have less than $0.5\%$, $0.09\%$ and $0.02\%$ likelihood of one-way daily trips taking longer that $100$ min, $200$ min and $400$ min, respectively. While neglecting these probabilities causes some error in the traffic values, the PCC (between flow data and model) saturates as function of the range limit, as shown in {\cg Supplementary Figs. 6c,d}.  In particular, at $100$ km or $100$ min the PCCs  are already close to their corresponding saturation values. Based on Fig. \ref{fig:3}b, this translates back to about $10^{-4}$ below which mobility flux values can be neglected.}

\section*{\large Acknowledgements}
We thank A.L. Barab\'asi, Z. N\'eda {\cg and H.T. Wang} for discussions. We also thank
R. Lychtenwalter for his computational help to speed up the betweenness algorithm and 
Sz. Horv\'at and M. Varga for a critical reading of the manuscript. This work 
was supported in part by US HDTRA 1-09-1-0039 (MER, YR and ZT),
the US NSF BCS-0826858, in part by grant FA9550-12-1-0405 from the
U.S. Air Force Office of Scientific Research and Defense Advanced
Research Projects Agency (ZT) and by a grant of the Romanian
CNCS-UEFISCDI, Project No. PN-II-RU-TE-2011-3-0121 (MER).  

\section*{\large Author Contributions} 
YR analyzed data, wrote simulation software, contributed tools, MER contributed tools and
designed methods, PW and MCG provided, prepared and analyzed data. ZT designed 
the research and wrote the paper.

\section*{\large Additional Information} 

{\bf Competing financial interests} The authors declare no competing financial interests.

\end{document}